# CONSENSUS Project: Identifying publicly acceptable policy implementations

Konstantinos Tserpes[1,2]

[1] Institute of Communication and Computer Systems,
National Technical University of Athens,
9, Heroon Polytechniou Str, 15773 Athens, Greece
`tserpes@mail.ntua.gr`
[2] Harokopio University of Athens
9, Omirou Str., 17778 Tavros, Greece
`tserpes@hua.gr`

**Abstract.** Even though it is unrealistic to expect citizens to pinpoint the policy implementation that they prefer from the set of alternatives, it is still possible to infer such information through an exercise of ranking the importance of policy objectives according to their opinion. Assuming that the mapping between policy options and objective evaluations is a priori known (through models and simulations), this can be achieved either implicitly through appropriate analysis of social media content related to the policy objective in question or explicitly through the direct feedback provided in the frame of a game. This document focuses on the presentation of a policy model, which reduces the policy to a multi-objective optimization problem and mitigates the shortcoming of the lack of social objective functions (public opinion models) with a black-box, games-for-crowds approach.

## 1 Introduction

A policy is meant to simultaneously attain multiple, often seemingly unrelated objectives. For instance, a policy regarding biofuels should take into consideration issues like fuel but also food prices. The policy context is a typically complex environment and as a result a policy implementation decision in order to meet one objective may trigger a set of changes in other fields, often with undesirable consequences or even conflicting results. A policy regarding the increase of state income may be implemented through the increase of VAT which in turn may result in recession growth and eventually possibly a decrease in the state income. As such, the policy maker needs to consider numerous factors that are affected by a certain policy implementation and pro-actively incorporate and evaluate as many objectives as possible.

The key challenge is to model existing real-world use-cases within the relevant policy-making context, and consequently employ measurable quantifiers in order to investigate how and whether preferable tradeoffs can be identified. Those quantifiers can be sought in multiple realms – such as analytical models, numerical simulations, statistical tools and even public opinion evaluators – in order to link the domain data to the set of objectives, and by that to reflect the expected success-rate of policies and their implementation.

Once the various alternative policy implementations are mapped to objective evaluations the policy makers can investigate the objective space and conclude to the policy implementation that best fits their preferences. Figure 1 depicts the objective space derived from the mapping between policy implementations and objectives.

|   | $Obj_1$ | $Obj_2$ | $Obj_3$ | $Obj_4$ | ... | $Obj_m$ |
|---|---|---|---|---|---|---|
| Policy Impl$_1$ |   |   |   |   |   |   |
| Policy Impl$_2$ |   |   |   |   |   |   |
| Policy Impl$_3$ |   |   |   |   |   |   |
| Policy Impl$_4$ |   |   |   |   |   |   |
| ... |   |   |   |   |   |   |
| Policy Impl$_n$ |   |   |   |   |   |   |

**Figure 1:** Policy implementation and objective evaluation mapping table.

In fact, once this mapping is done then the problem of identifying the policy implementation that meets the objective in the best possible way is reduced to the identification of the Pareto-optimal solutions, i.e. those solutions that there are superior to any other solution from the set. At this point it is worth noting that the Pareto-optimal solutions are often multiple and further narrowing down the set to a single policy also depends on a certain preference or disposition towards one or another objective. For instance, in Figure 2, points A, B and C are equally considered optimal as there are no points that dominate them in both axes simultaneously: point A is the highest in the y axis, point C is the highest in the y axis and for point B there is no single point that has higher values in axes x and y simultaneously.

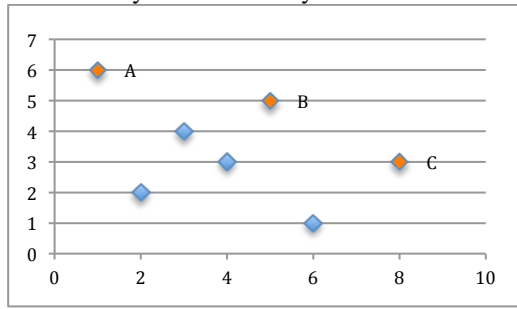

**Figure 2:** Illustration of a 2-objective space (example). The Pareto frontier is formed from points A, B and C. Optimization in this example is referring to maximization.

In the abovementioned, simplified example, the policy maker could pick any of these three points in order to maximize the yield of the policy implementation, however it is entirely up to her/him to select one. If s/he would like to put more emphasis on the y axis, s/he would pick point C, etc.

As such, the suggested model reduces a policy making problem to a strictly mathematical problem. However, it is meant to be used as a model that provides insights to the user rather than replacing the decision making process. Pareto optimality is not meant to replace any system of social choice which are governed by social rules, although it is proven that any such system will eventually converge to Pareto efficient, but inequitable, distributions [1].

Having stated that, it is regarded imperative to engage citizens in the decision making process an endeavor highlighted in all state of the art analyses such as the Code of Good Practice on Civil Participation in the Decision-Making Process [2]. We therefore need to seek the citizens' involvement in policy making since their input can potentially become highly valuable in various stages, from gathering the necessary data, through formulating public opinion as one of the objectives in the model, to eventually playing the role of exploring the attained tradeoffs and contributing to their weighing. A positive by-product of this process is the education of the citizens in matters of policy implementations, also contributing to the transparency in policy making.

The rest of this document converses about a design, implementation and experiment conducted in the frame of the Consensus project[1] with the intention to incorporate the public opinion to the policy making process. In particular, the following section presents the approach that was followed in order to achieve the abovementioned goal. Section 3, provides the details and highlights of the evaluation plan that was executed with the help of external users, and Section 4 presents the related work in the fields that inspired, influenced and are comparable to this work. Finally, Section 5 closes this document with the main conclusions made out of this work.

## 2 Approach

The goal of engaging citizens in the policy making process needs to stem from particular requirements rather than blindly aim in their involvement. As such the design and implementation of the software tool and underlying model in the Consensus project derived from two specific important requirements:

- The education of citizens regarding the consequences of certain policy implementation options
- The harvesting of user preferences so as to include the public opinion as an objective in the policy making

The overarching concept of the software system and model that can meet these two requirements, involves the direct evaluation of policy implementations by citizens so as for them to acquire further details about the objective evaluations when selecting one policy and secondly help policy makers see what people prefer in terms of policies.

The main challenges in this endeavor are the following:

- The typical problem of user engagement [3] with information systems which may lead to digital exclusion [4], a situation conflicting with the initial goal.

---

[1] http://www.consensus-project.eu

- The fact that the citizens cannot evaluate the policy implementations without proper guidance: a) because policies often contain details unknown to the citizens (technicalities) and b) because the policy implementations evaluation can lead to a huge number of options which the citizens cannot practically investigate.

These challenges had a deep impact in the solution design. First and foremost, the user incentivation issue which is promoted though gamification techniques [5]. In particular visualizations and assistants were employed in order to introduce citizens in the policy context and guide them through the process. At the same time, points and badges were assigned to actions that inferred a positive learning process and a scoreboard was maintained in order to enhance the competitive nature of the tool.

This system was implemented and made available to the public as a web-based platform that was named: "Consensus Game"[2]. Note that disregarding the name, the system is not considered to be a web game not even a game of any form. It merely employs concepts commonly found in games with the purpose to be appealing to users and ensure that the process promotes is a certain goal, in this case, education about policies and policy making.

The answer to the second challenge is deeply rooted in the flow of logic of the Consensus Game (the term "Game" is used interchangeably in this document) and linked to the policy model outlined in Section 1. The details are better described in what follows.

### 2.1 Consensus Game

The main process served in the Consensus Game dictates that the citizens are called to evaluate the objectives, rather than the policies. The objectives are often high level concepts (e.g. cost of food, $CO_2$ emissions, road safety, etc) which are closer to the layman's understanding. This approach is dictated by the first challenge introduced in the previous section.

The main idea is to have people explain which objective is more important for them while appreciating that not all options are feasible under realistic conditions. The benefit from this activity is twofold:
a) the citizens provide direct feedback to the policy makers regarding the public opinion's priorities.
b) this information can be integrated in the decision making process and further narrow down the Pareto frontier options.

By ranking all of the objectives (or prioritizing them) practically results in the filtering of the underlying policy implementation alternatives. That is, by placing an objective in the top of the priority list the citizen is implying that the solutions that achieve this objective better are preferred than the rest (Figure 3). A side effect is that this process may result in excluding Pareto efficient solutions, however, in an adequately large set of alternatives this will merely result in narrowing down the Pareto-optimal solutions. The policy maker can then use this information in order to under-stand the public opinion preferences and integrate it in the policy model and the policy making process, either as an extra objective or by directly focusing on the top alternatives from the list.

---

[2] Consensus Game is available at: http://platform.consensus-project.eu/consensus/

|             | Obj1 | Obj2 | Obj3 |     |             | Obj2 | Obj3 | Obj1 |
|-------------|------|------|------|-----|-------------|------|------|------|
| **Policy Alt1** | 8 | 0 | 3 |     | **Policy Alt3** | 10 | 8 | 3 |
| **Policy Alt2** | 5 | 6 | 9 |     | **Policy Alt2** | 6 | 9 | 5 |
| **Policy Alt3** | 3 | 10 | 8 |    | Policy Alt4 | 4 | 1 | 5 |
| Policy Alt4 | 5 | 4 | 1 |         | **Policy Alt1** | 0 | 3 | 8 |

**Figure 3:** Mapping on the left demonstrates that the policy alternatives #1,#2 and #3 are Pareto-optimal and should considered as equally good options. Mapping on the right shows a different prioritization (Obj2,Obj3,Obj1) in which policy alternative #3 is preferred (social choice)

The educational part in this process is achieved by introducing a subsequent step after allowing the users to define objective prioritizations. This step involves the presentation of policy implementations that link back to the selected objective prioritization. I.e., after the citizens create a prioritized list of objectives, the system presents them those alternatives that meet the objectives in this priority in the best possible way. The details can be provided with visualizations in relation to the objectives' evaluation as well as with links and small and simple descriptions. The idea now is to introduce the users to the policy options allowing them to toy with the objective priority.

With this concept in mind an evaluation session was executed, in which users participated online, registering with the Consensus Game and selecting one of two available policy scenarios: one related to biofuel and one related to transportation (road infrastructure funding). Then they were presented with the objectives and appropriate visualizations where they could pick priorities for the set of objectives, as if they were policy makers (but at a different level). After they make their selection they were presented with the set of policies that lead to this prioritization (for usability purposes max 3 are presented: an optimal and two inferior, all randomly picked). The citizens were able to investigate the details around these policies with short texts and links to examples and even legislation documents for the more advanced users. They were also able to see the degree to which each policy fulfills the complete set of objectives (based on the prioritization they selected).

In what follows the details of this evaluation session are presented.

## 3 Evaluation

The evaluation pilot of the Consensus Game was planned and executed in a 2-month time span. The following subsections provide a brief account of the pilot plan and the main outcomes from its execution.

## 3.1 Evaluation plan

The end users of the Consensus Game are constituted largely by anyone considered a citizen interested to policy making. In particular the target was a broad audience included in personal and business social networks.

The main evaluation method was based on the analysis of the answers that the users provided in the questionnaires. As such, a web page that contained the complete information needed in order to play the game was created. The link to this page was sent through emails, social media posts along with a link to a questionnaire that the citizens had to answer to complete the evaluation.

A secondary method included the analysis of the server log while the players were using the platform. This gave us hints about how people perceive the provided information by monitoring the user interaction with the page.

The plan of the evaluation session for the Game included various steps, in particular:

For a period of two months the links with the invitation, instructions, platform page (http://platform.consensus-project.eu/consensus) and questionnaires were sent to all possible networks to which the consortium members, partners and the project itself had access. The process involved the gradual extension of the networks to which we made the pilot Game available with the first step including a test run. In particular the test run included the sending of the invitation within the consortium organizations from which immediate feedback could be collected and quickly incorporated before sending it out to unknown recipients. Two days were allowed for this phase and after minor glitches were fixed, the second phase was initiated in which the invitations were sent to a large amount of users (>=300) through the project's communication channels, including mailing lists and social media accounts. After another 2 days, the invitation was sent to selected mailing lists. Finally, the Game was showcased in the relevant conference "Gaminomics"[3] that took place in London on the 11th of June in which oral feedback was collected and incorporated in the evaluation.

Eventually, the anticipated number of at least 100 users to start the game, was achieved through this method however not all users answered the questionnaires or provided feedback in any way (see Table 1).

**Table 1.** Consensus Game evaluation log

| Target Group | Dissemination means | Evaluation Feedback | Audience Size | Participated | Received feedback from |
|---|---|---|---|---|---|
| Project meeting participants | Focus group | Discussion | 20 | 20 | 20 |
| Project Partners' groups | Email | Online questionnaires/Oral feedback | 40 | 40 | 30 |
| Project dissemination tools targets | Email, LinkedIn, Twitter, banners in website, etc | Online questionnaires/Oral feedback | >300 | 60 | 16 |
| Related projects | Email | Online questionnaires/Oral feedback | 50 | 0 | 0 |
| Gaminomics Conference | Conference presentation | Discussion | 100 | 10 | 5 |

---

[3] http://gaminomics.com

## 3.2 Evaluation Outcomes

Based on the plan laid out in the previous subsection the evaluation of the Consensus Game took place in multiple steps and through multiple means for dissemination and evaluation feedback. A total 33 questionnaires were fully filled out and about 15 people provided oral feedback. The game was played by 53 people who played a total of 241 times. The degree distribution is skewed towards the left side of the chart can be seen in Figure 4.

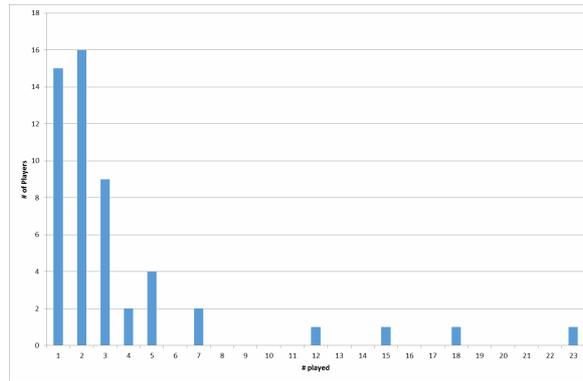

**Figure 4:** Participation degree distribution

The degree distribution highlights a few interesting conclusions:
- Most of the people played a few times, which implies that the incentive was not strong enough for them to be engaged.
- The biggest portion of the players played 2 and 3 times which is probably indicative of the fact that they tried to understand the Game.
- The players that played only once were either not interested all along or dissuaded by the result
- 4 players engaged heavily with the Game apparently attempting to score high in the scoreboard (see Figure 5)

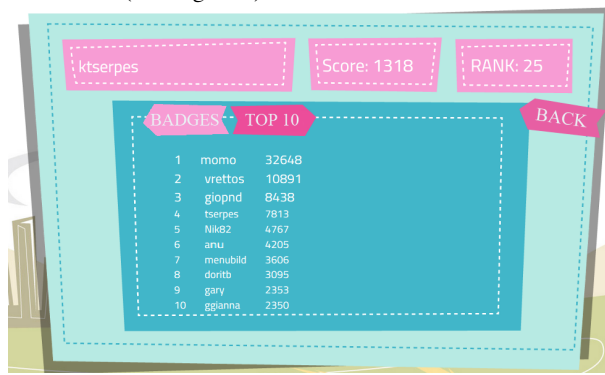

**Figure 5:** Game top 10 scoreboard as of 23/06/2015

Of the 241 game sessions, the 132 were for the biofuel scenario, while the rest 109 were for the transportation scenario.

Further analysis of the results show that the most popular objective prioritization for the two scenarios was:

- Biofuels: 2112, i.e. CO2 Emissions and Cost of Food as set as No1 priorities and Forest Land and Biodiversity are set as No2 priorities.
- Transportation: 322413, i.e. User convenience (#1), Noise and Accident Cost (equally to #2), Levels of Service Charge and Alternative Routes and Modes (equally to #3) and Air pollution is prioritized as low as #4.

Furthermore the evaluation of questionnaires and face-to-face feedback revealed that the current version of the Game manages to achieve the first requirement completely but it does not address the second well. Players need to prioritize a set of objectives (Figure 6) and the results from this step are particularly useful in the narrowing down of the set of optimal solutions for the policy makers.

Even though objective prioritization and all the constraints became instantly clear to the users, the introduction of a second step in which policy implementations are presented and citizens are asked to identify the optimal one, made the process complex. Exactly because policies and policy implementations are complex by nature, it was difficult to meet the education part well.

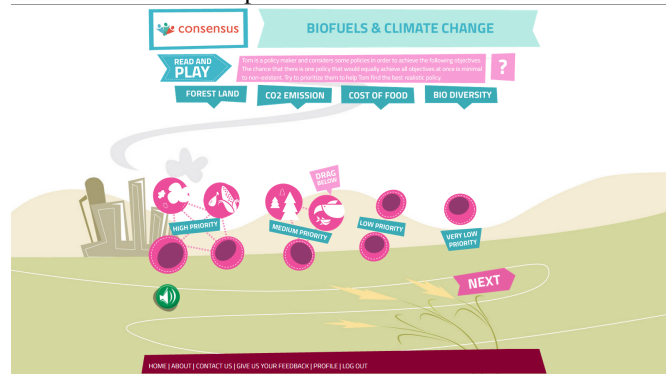

**Figure 6:** First part of the Game, objective prioritization (preferences elicitation)

In order to mitigate this problem we resorted to storytelling and assistants. During the evaluation phase we observed that this helped significantly, but only to those players who were meticulous and patient to use the guides. Since this is not often the case, the measure's success was mediocre.

Another shortcoming that related to the policy education part was the fact that the task of finding the optimal solution turned to be a visual recognition challenge. The players had to go through charts indicating the degree to which each presented policy met the objective prioritization goal (Figure 7) and decide which one is better. This task is clearly difficult and probably not interesting at all, at least not in the way presented.

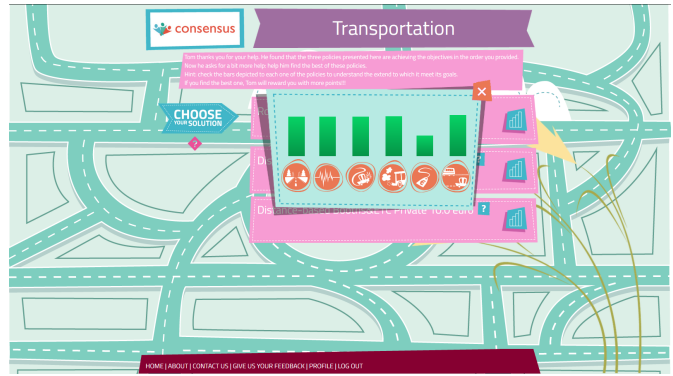

**Figure 7:** Second part of the Game, policy selection (education part)

## 4 Related Work

This black box approach for modelling public acceptability and opinion aims at directly relaying the issue of policy implementation acceptance to the citizens and have them evaluate a certain set of objective values in a specified form. This black box approach models the public acceptance objective function in a consistent way.

Tools like Ushahidi [6], crowdflower [7] and crowdsource.com and IdeaScale.com and Mechanical Turk (mturk.com) for the US, can be used in order to deploy and disseminate tasks to crowds as well as collect data. Common social media platforms like Twitter and Facebook can be also used for the same purpose. However, such (raw) data obtained from Social Web feeds often contain variable amounts of "noise", misinformation and bias (which can get further "amplified" through the viral nature of social media) and will usually require some advanced forms of filtering and verification by both machine-based algorithms and human experts before becoming reliable enough for use in decision-making tasks. WSARE (What's Strange About Recent Events)-type algorithms [8] and platforms such as SwiftRiver [9] (open source, provided by Ushahidi) can prove helpful in trying to filter the Social Web "firehose".

Disregarding the implementation platform and network, one of the most difficult challenges of crowdsourcing is how to draw and retain users to the crowdsourcing system. One strategy that is widely used is combining crowdsouring with gamification. According to Von Ahn et. al.[10] a crowdsourcing game is: a) Fun and engaging; b) Includes a task that can only be completed by humans; and c) Has a goal that is hidden from the player.

Prominent examples of such games are the ESP Game [11] where two players receive an image as input and need to "agree" on as many tags as possible that describe the given image and GuessWho [12], in which enterprise employees enter knowledge about their peers to enrich the organizational social network. Other notable implementations are TagATune [13], Peekaboom [14],Verbosity [15], Curator [16], PageHunt [17] and Collabio [18].

Other interesting approaches are investigated in SocIoS[4] and +Spaces[5] projects. In the first, the users were given a script of a TV commercial and they were asked to submit their photos if they believed that they should take a role in the commercial. Then the crowds were requested to rank those participants that they felt were the right ones for the role. A similar exercise was done for location scouting (finding the location for the scouting). The users were rewarded with fun points and badges. This was a sort of explicit collaboration of the users with the system, i.e. the users actively contribute content.

+Spaces is a case in which the users were implicitly collaborating with the system, i.e. user actions are recorded and processed without their direct contribution (but with their consent). In that case, social media and virtual worlds were leveraged so as to simulate a policy context and users, through their interaction with the system, were generating feedback to the policy maker. Another such example of implicit collaboration is presented in [19] in which the game initiators capture implicit behaviour traces from online crowd workers and use them to predict outcome measures such as quality, errors, and likelihood of cheating.

All the abovementioned use cases fall under the broad category of Games With A Purpose (GWAP) [10], propose that using computer games can gather human players and solve open problems as a side effect of playing.

Other similar, notable approaches in using gamification for policy making include: SimCityEDU: Pollution Challenge[6] and IBM CityOne[7]. These games are based on gaming platforms in which they have developed virtual worlds and economies. They consider that the user will spend considerable time in the platform and even though their educational capacity is undeniable, they do not meet the second requirement posed in the proposed policy model, i.e., the elicitation of user preferences and their integration in the policy making process.

## 5 Conclusions

A policy model based on reducing a policy making process to a multi-objective optimization problem is introduced. In such a mathematical formulation, the identification of optimal solutions is feasible and various visualization and decision support tools can be used to assist the policy maker explore the objective space. However, there are no models for the social acceptability of the policy alternatives. In order to accommodate this shortcoming and enable the integration of the public preferences to the decision making process of our policy model, we propose a process and a system that revolves around two related requirements:

- Preference elicitation for the incorporation of the public opinion in the policy making process
- Citizen education regarding the policies and policy making process

---

[4] http://www.sociosproject.eu  
[5] http://www.positivespaces.eu  
[6] https://www.glasslabgames.org/games/SC  
[7] http://www-01.ibm.com/software/solutions/soa/innov8/cityone/

The evaluation of the tool revealed that the current version of the Game manages to achieve the first requirement completely but it does not address the second well.

Among the things that will definitely improve the Game's effectiveness is the incorporation of "feedback". The users need to understand why a certain selection leads them to a specific result. In order to deal with that, we plan to introduce real world examples and visually correlate them with their decisions and possible options. Furthermore, we plan to use short videos and tutorials to smoothly introduce the user to the context.

To sum up, the current version of the Game was a successful experiment to show whether citizens can be engaged in policy-making processes by employing gamification concepts. In terms of exploitation, one needs to consider that these solutions are open-ended and creativity plays a huge importance. Therefore there is no reliable way to claim that there is a specific implementation that works. Consensus Game can only demonstrate that gamification is an approach that can indeed trigger the interest of citizens.

# 6 Acknowledgement

This work has been supported by the Consensus project (http://www.consensus-project.eu) and has been partly funded by the EU Seventh Framework Programme, theme ICT-2013.5.4: ICT for Governance and Policy Modelling under Contract No. 611688.
The author would also like to acknowledge the work of the Athens Technology Center team and in particular Mr. Nikos Dimakopoulos, Leonidas Kallipolitis and Anna Triantafyllou who developed the tool.